\documentclass[twocolumn,aps,prd,floatfix]{revtex4-2}
\usepackage[T1]{fontenc}
\usepackage{amsmath,amssymb,amscd}
\usepackage{graphicx}
\usepackage[leftcaption]{sidecap}
\usepackage[colorlinks,allcolors=blue]{hyperref}

\usepackage[most]{tcolorbox}
\newtcbox{\othermathbox}[1][]{nobeforeafter, math upper, tcbox raise base, enhanced, rounded corners, colback=black!5, colframe=black, left=0.7em, top=0.4em, right=0.7em, bottom=0.5em}
\usepackage{empheq}

\usepackage{color}
\usepackage{enumitem}
\usepackage{stmaryrd}
\usepackage{mathrsfs}
\usepackage{amsfonts}
\usepackage{amssymb,scalerel,stackengine}
\newcommand\beq{\begin{equation}}
\newcommand\ee{\end{equation}}
\newcommand\cO{{\cal O}}

\newcommand\dd{\text{d}}

\newcommand\der{\partial}
\newcommand\ie{{\it i.e.}\ }
\newcommand\eg{{\it e.g.}\ }

\def\phii{\varphi}

\def\hc{{\hat c}}

\def\de{\delta}

\def\pd{\partial}

\def\eps{\epsilon}
\def\veps{\varepsilon}
\def\cC{\mathcal{C}}

\def\cA{\mathcal{A}}
\def\cC{\mathcal{C}}

\def\cF{\mathcal{F}}

\def\cO{\mathcal{O}}

\def\bm{\boldsymbol{m}}
\def\bn{\boldsymbol{n}}
\def\bp{\boldsymbol{p}}
\def\bv{\boldsymbol{v}}
\def\bx{\boldsymbol{x}}
\def\by{\boldsymbol{y}}

\def\bR{\boldsymbol{R}}

\def\be{\boldsymbol{e}}
\def\bJ{\boldsymbol{J}}

\def\bx{\boldsymbol{x}}

\def\bOmega{\boldsymbol{\Omega}}
\def\bth{\boldsymbol{\theta}}

\def\th{\theta}
\def\veps{\varepsilon}
\def\tlA{\widetilde{A}}

\def\hc{{\hat c}}

\def\hc{\text{h.c.}}

\begin{document}

\title{Orientation Memory of Magnetic Dipoles}

\author{Blagoje Oblak}
\affiliation{CPHT, CNRS, Ecole Polytechnique, IP Paris, F-91128 Palaiseau, France.}
\author{Ali Seraj}
\affiliation{School of Mathematical Sciences, Queen Mary University of London, Mile End Road, E1 4NS, United Kingdom.}

\begin{abstract}
We study the precession caused by electromagnetic radiation on a magnetic dipole located far from the source. As we show, this entails a net rotation of the dipole in the plane orthogonal to the direction of wave propagation, providing an electromagnetic analogue of gyroscopic gravitational memory. Like its gravitational cousin, the precession rate falls off with the square of the distance to the source, and is related to electric-magnetic duality and optical helicity on the celestial sphere. We use a multipolar expansion to compute the memory effect due to localized sources such as moving point charges, and highlight its occurrence in setups that break parity symmetry.
\end{abstract}

\maketitle

\section{Introduction}

The passage of a burst of gravitational waves through a detector typically leads to persistent effects, generally known as \textit{gravitational memory} \cite{Zeldovich:1974gvh,Braginsky:1985vlg,braginsky1987gravitational,Christodoulou:1991cr}. Such phenomena have received widespread attention in recent years. Indeed, the seminal detection of gravitational waves \cite{LIGOScientific:2016aoc} and the growing number of subsequent observations provide encouraging prospects for the detection of memory in the near future \cite{Lasky:2016knh,Hubner:2019sly,Hubner:2021amk,Boersma:2020gxx,Grant:2022bla,Ghosh:2023rbe,Gasparotto:2023fcg}, while the rich interplay between memory effects, asymptotic symmetries and soft theorems in quantum gravity \cite{Strominger:2014pwa,Strominger:2017zoo} makes them crucial from a theoretical standpoint. This has led to numerous investigations of memory effects, corresponding observables, and their relation with gravitational charges: see \eg the sample \cite{Favata:2010zu,Bieri:2013ada,Pasterski:2015tva,Garfinkle:2017fre,Nichols:2017rqr,Nichols:2018qac,Satishchandran:2019pyc,Bieri:2020pee,Bieri:2020zki,Tahura:2020vsa,Tahura:2021hbk,Seraj:2021qja,Bernard:2022noq,Heisenberg:2023prj} and references therein.

As is turns out, memory effects are not limited to general relativity and occur quite generally in any gauge theory. It was indeed shown in \cite{Bieri:2013hqa} that electromagnetic waves cause a net change in the velocity of test charges near null infinity. This is the ``kick'' memory effect of electrodynamics, the simplest gauge-theoretic counterpart of the displacement memory produced by gravitational radiation \cite{Zeldovich:1974gvh,Braginsky:1985vlg,braginsky1987gravitational,Christodoulou:1991cr}. Analogous phenomena were later identified in other gauge theories \cite{Pate:2017vwa,Ball:2018prg,Jokela:2019apz} and massless field theories more generally \cite{Satishchandran:2019pyc}. In hindsight, this ubiquitousness of memory effects is no surprise, since soft theorems and asymptotic symmetries similarly occur in gauge theories other than gravity \cite{He:2014cra,Campiglia:2015qka,Strominger:2015bla,Conde:2016csj,Gabai:2016kuf}.

A key aspect of memory effects is their dependence on the distance $r$ between source and detector. This entails a ``hierarchy'' of leading and subleading observables, according to the rate of their decay at large $r$, but it also implies differences in the way leading and subleading effects are meant to be measured. For instance, the leading memory effect in general relativity is a change of distance between test masses, behaving as $1/r$ at large $r$ \cite{Zeldovich:1974gvh,Braginsky:1985vlg,braginsky1987gravitational,Christodoulou:1991cr}. The same $1/r$ fall-off is satisfied by the aforementioned kick memory of electrodynamics \cite{Bieri:2013hqa}. At subleading order $1/r^2$, memory effects encode information about subleading soft theorems and various extended asymptotic symmetries  \cite{Pasterski:2015tva,Nichols:2017rqr,Nichols:2018qac,Grant:2021hga,Seraj:2022qqj}. For our purposes, the most relevant example in that context is that of refs.\ \cite{Seraj:2021rxd,Seraj:2022qyt}, which describe a \textit{gyroscopic memory effect}: a net rotation of a spinning gyroscope in the ``transverse plane'' orthogonal to the direction of gravitational wave propagation.

The present work is devoted to an electromagnetic analogue of gyroscopic memory. It consists of a persistent rotation caused by electromagnetic radiation on the orientation of a distant magnetic dipole. This observable, which we dub ``gyromagnetic memory,'' is remarkably similar to its gyroscopic gravitational cousin \cite{Seraj:2021rxd,Seraj:2022qyt}. Indeed, both effects are subleading in that they decay as $1/r^2$. Furthermore, gravitational gyroscopic memory contains two terms: one that is linear in the metric perturbation and coincides with the spin memory effect \cite{Pasterski:2015tva}, and a second, nonlinear part related to gravitational electric-magnetic duality and the helicity of gravity waves. The same structure turns out to arise in electrodynamics, despite one's naive expectation that no nonlinear term should arise in Maxwell's linear theory \footnote{The appearance of a nonlinear term in electrodynamics is technically due to the solution of the spin precession equation, which involves the expansion of a time-ordered exponential in the magnetic field.}. In particular, the nonlinear term is again related to electric-magnetic duality and the optical helicity of radiation. Similar quantities also occur in \cite{Maleknejad:2023nyh,Liu:2023qtr}, which appeared while we were finalizing this work.

An obvious advantage of the electromagnetic setup compared to its gravitational version is its simplicity: one can compute, with minimal effort, the radiative data and memory caused by a given source. (This should be contrasted with the gravitational case, where the extraction of radiative data from dynamical sources involves intricate numerical or perturbative frameworks \cite{Blanchet:2013haa,Buonanno:1998gg,Porto:2016pyg,Barack:2018yvs,Bern:2019nnu}.) Accordingly, we eventually study the gyromagnetic precession produced by nonrelativistic moving point charges and find that it crucially requires a breaking of parity symmetry. This occurs for instance in the simple case of a rotating point charge, suggesting that a similar gravitational gyroscopic precession occurs for inspiralling binary black holes; these will be studied in a separate work.

The paper is organized as follows. In section \ref{sepre}, we show how electromagnetic radiation gives rise to the precession of a magnetic dipole near null infinity. Section \ref{sedual} is then devoted to the relation between the resulting memory effect, electric-magnetic duality and optical helicity. Finally, in section \ref{sesource} we compute the gyromagnetic precession and memory produced at null infinity by moving point charges in the bulk; this involves in particular a multipolar, nonrelativistic expansion, also used in gravitational computations of the same kind.

\medskip
\noindent\textbf{Notation.} We use Gaussian units and set $c=1$ throughout, except at the very end of section \ref{sesource}. Vectors are denoted by bold letters, \eg the position $\bx=x^i\partial_i$ and the radial unit vector $\bn\equiv\bx/|\bx|=n^i\partial_i$. In addition, we interchangeably use $\bn$ and $\bth$ to represent a point on a unit (celestial) sphere. We also define $\Delta f(u)\equiv f(u)-f(-\infty)$ and $\Delta f\equiv\lim_{u\to\infty}\Delta f(u)$ for any time-dependent function $f(u)$ whose derivatives vanish at past and future infinities. Finally, in section \ref{sesource}, we will rely on the multi-index notation $X_L=X_{i_1i_2\cdots i_\ell}$ to construct symmetric trace-free (STF) harmonics $\hat n_L\equiv n_{\langle L\rangle}$, where angle brackets denote the STF part of a tensor. 

\section{Gyromagnetic precession \texorpdfstring{\\and memory}{}}
\label{sepre}

Here we use the asymptotic behaviour of electrodynamics near null infinity to predict the precession rate of a magnetic dipole located far away from a source of electromagnetic waves, at leading order in the inverse distance from the source. For radiation bursts that are compactly supported in time, this leads to a net gyromagnetic memory, which we compute.

\medskip
\noindent\textbf{Asymptotic electromagnetic field.} Consider four-dimensional Minkowski spacetime with inertial coordinates $x^{\mu}=(t,x^i)$, $i=1,2,3$, and define retarded Eddington-Finkelstein (Bondi) coordinates by $r\equiv\sqrt{x^ix^i}$ and $u\equiv t-r$. Let also $\th^B$ ($B=1,2$) be local coordinates on a (future) unit celestial sphere whose metric is $\gamma_{BC}(\bth)\,\dd\th^B\,\dd\th^C$. For later reference, introduce a time-independent orthonormal dyad $E_a{}^B(\bth)$ on $S^2$ such that $E_a{}^B E_b{}^C\gamma_{BC}=\de_{ab}$, with frame indices $a,b\in\{1,2\}$. One can then define a local Cartesian frame
\begin{align}
\label{frame}
\be_{r}&=n^i\pd_i\,,
\qquad
\be_{a}=r E_a{}^B\frac{\pd n^i}{\pd\theta^B}\pd_i
\end{align}
with $n^i\equiv x^i/r$. This frame will eventually be used to write the components of a magnetic dipole.

Now let $J_{\mu}(t,\bx)$ be some localized, generally time-dependent, electric current density in the bulk of Minkowski spacetime. It produces an electromagnetic field $\cA_{\mu}$ which, in Lorenz gauge $\der_{\mu}\cA^{\mu}=0$, satisfies the Maxwell equation $\Box \cA_\mu=-4\pi J_{\mu}$. The corresponding causal solution  is
\begin{align}
\label{solMax}
    \cA_\mu(t,\bx)
    &=
    \int\!\dd^3\by\int\limits_{-\infty}^t\!\!\dd t'\,\frac{\delta(t'-t+|\bx-\by|)}{|\bx-\by|}J_\mu(t',\by).
\end{align}
The assumption of a compact source allows us to expand $|\bx-\by|=r-\bn\cdot\by+\cO(1/r)$ at large $r$, so that \eqref{solMax} gives access to the electromagnetic field near future null infinity, \ie in the limit $r\to\infty$ with finite retarded time $u=t-r$. One finds indeed
\begin{align}
\label{aasp}
    \cA_\mu(t,\bx)&=\frac{1}{r}A_\mu(u,\bn)+\cO(r^{-2})
\end{align}
where $A_\mu(u,\bn)$ is a smooth function on null infinity, determined by the source according to
\begin{align}
\label{arad}
    A_\mu(u,\bn)&=
    \int \dd^3\by\,J_\mu(u+\bn\cdot\by,\by)\,.
\end{align}
The Lorenz gauge condition then becomes equivalent, at leading order, to the conservation equation $\der_{\mu}J^{\mu}=0$.

The leading components of the electric and magnetic fields can immediately be deduced from \eqref{aasp} in terms of the asymptotic data $A_\mu(u,\bn)$. In particular, the magnetic field $B^i=\frac{1}{2}\eps^{ijk}\cF_{jk}$ expressed in the local frame \eqref{frame} is given by
\begin{equation}
\label{fexp}
\cF_{ra}
=
-\frac{1}{r}\dot A_a
+\cO(1/r^2),
\quad
\cF_{ab}
=
\frac{1}{r^2}F_{ab}+\cO(1/r^3)
\end{equation}
where $A_a\equiv E_{a}{}^{B}\tfrac{\pd n^i}{\pd\th^B}A_i$ and $F_{ab}\equiv\pd_a A_b-\pd_b A_a$ with $\pd_a\equiv E_a{}^A\frac{\pd}{\pd \th^A}$. Note the different fall-offs of radial and angular components of  the field strength; this will affect the large-$r$ expansion of the precession equation, to which we now turn.

\medskip
\noindent\textbf{Precession and memory.} 
Consider a static magnetic dipole $M_i(t)$ in Minkowski spacetime, with $i=1,2,3$ an index in any local Cartesian frame. When the dipole is exposed to a magnetic field $\cF_{ij}$, its evolution equation reads
\begin{align}
\label{dipev}
\dot M^{i}=k\,\cF_{ij}\,M^{j}.
\end{align}
where $k$ is the dipole's gyromagnetic ratio and $\cF_{ij}$ is evaluated at the dipole's location. (Note: since the frame is Cartesian, indices may be raised and lowered at will.) The general solution of eq.~\eqref{dipev} is a time-ordered matrix exponential
\beq
\label{orexp}
M^i(t)
=
\underbrace{T\exp\left(k\int\limits_{-\infty}^t\dd u\,\big[\cF_{ij}(u)\big]\right)}_{\displaystyle\equiv W_{ij}(t)}\cdot\,M^j(-\infty),
\ee
where we formally take the initial point to be $t_0=-\infty$, implicitly assuming that the field $\cF_{ij}(u)$ is compactly supported in time.

Now let $\cF_{ij}$ be produced by a distant dynamical source, and express the precession equation \eqref{dipev} and its solution \eqref{orexp} at large $r$ in terms of the asymptotic data \eqref{aasp}. One can indeed use the field strength \eqref{fexp} to expand the rotation matrix $W_{ij}$ in \eqref{orexp} as
\begin{align}
\label{wra}
    W_{ra}&=\frac{k}{r}\int_{-\infty}^t\dd u\,\cF_{ra}= -\frac{k}{r}\Delta A_a(t)+\cO(r^{-2})\\
    W_{ab}&=\de_{ab}+\frac{k}{r^2}\bigg[\int_{-\infty}^t \dd u\, F_{ab}\nonumber\\
    \label{wab}
    &\quad-k \int_{-\infty}^t \dd u\int_{-\infty}^{u}\dd v \dot{A}_a(u)\dot{A}_b(v)\bigg]+\cO(r^{-3}),
\end{align}
where we chose the Cartesian frame \eqref{frame} to distinguish radial and angular directions and assumed that the dipole is located at a fixed position in Cartesian coordinates $x^i$. In particular, the precession rate in the transverse plane is encoded in the antisymmetric part of the integrand of eq.~\eqref{wab}, \ie $\Omega \equiv \frac{1}{2}\eps^{ab}\dot{W}_{ab}$, which in terms of the radiative data reads
\begin{align}
\label{omm}
    \Omega(u)
    &=
    \frac{k}{r^2} \left(D_a \widetilde{A}^a(u) -\frac{k}{2} \dot{A}^a(u)\Delta\widetilde{A}_a(u)  \right),
\end{align}
where $\widetilde{A}_a\equiv\eps_{a}{}^b A_b$ is the dual asymptotic data corresponding to \eqref{aasp} and $\Delta \tlA_a(u)\equiv\tlA_a(u)-\tlA_a(-\infty)$.

The precession rate \eqref{omm} is remarkably similar to its gravitational cousin. Indeed, the precession rate of a gyroscopic subjected to gravitational radiation is \cite{Seraj:2021rxd,Seraj:2022qyt}
\begin{align}
\label{s3b}
    \Omega_{\text{grav}}
    &=
    \frac{1}{4r^2}
    \left(D_AD_B\widetilde C{}^{AB}-\frac12\dot{C}^{AB}(u)\widetilde{C}_{AB}\right),
\end{align}
where $\widetilde C{}_{AB}\equiv\epsilon^C{}_AC_{BC}$ is the dual shear tensor at null infinity and $\dot{C}_{AB}$ is the news tensor. This is exactly the form of eq.~\eqref{omm}, with $\widetilde{A}$ replaced by dual shear. (All results being covariant on $S^2$, one is free to exchange dyad indices $a,b$ for coordinate indices $A,B$.) The only key difference is the presence of the coupling constant $k$ in eq.~\eqref{omm}, which is absent in the gravitational case \eqref{s3b} owing to the equivalence principle.

It is immediate to deduce the net rotation angle in the transverse plane following a burst of electromagnetic radiation: it is an integral $\Phi=\int_{-\infty}^\infty\dd u\,\Omega(u)$ with $\Omega$ given by eq.~\eqref{omm}. This angle is the \textit{gyromagnetic memory}:
\begin{align}
\label{memory}
    \Phi
    &=
    \frac{k}{r^2}
    \int\limits_{-\infty}^\infty \dd u\left(D_B\widetilde{A}^B(u) -\frac{k}{2} \dot{A}^B(u)\Delta \widetilde{A}_B(u)  \right),
\end{align}
where $A_B=E^a_{\;B}\,A_a$ and $D_B$ is the covariant derivative determined by the metric $\gamma_{BC}$ on $S^2$. Again, the similarity between this expression and gravitational gyroscopic memory \cite{Seraj:2021rxd,Seraj:2022qyt} is striking: the gravitational result is just obtained by integrating eq.~\eqref{s3b} over time.

\newpage
\section{Magnetic field,\texorpdfstring{\\duality and helicity}{}}
\label{sedual}

Here we study the two terms of the gyromagnetic memory effect \eqref{memory}. We start with the linear piece, which is nothing but the (time integral of the) radial magnetic field. Then we turn to the nonlinear term and show that it can be interpreted in two equivalent ways: either as a generator of local electric-magnetic duality on a celestial sphere, or as a measure of the difference between the numbers of left- and right-handed photons crossing a given point on the celestial sphere. As before, analogous interpretations hold in the gravitational version of the setup \cite{Seraj:2021rxd,Seraj:2022qyt}.

\medskip
\noindent\textbf{Linear term.} The term $2D_B\widetilde{A}{}^B=\epsilon^{AB}F_{AB}$ in eq.~\eqref{omm} is manifestly the radial magnetic field at null infinity. A less trivial fact is that $D_B\widetilde{A}{}^B$ may also be seen as a boundary current for dual large gauge transformations \cite{Strominger:2015bla}. Thus, it is analogous to the dual mass aspect \cite{Godazgar:2018qpq,Godazgar:2018dvh,Kol:2019nkc,Oliveri:2020xls,Kol:2020ucd,Freidel:2021qpz} that appears in the gravitational case \cite{Seraj:2021rxd,Seraj:2022qyt}. Note that this can be extrapolated to magnetic monopoles: in that case, $D_B\widetilde{A}{}^B$ would be replaced by a nonzero constant, leading to a constant precession rate \eqref{omm} on top of the radiative contribution. The gravitational analogue of this situation would be a gyroscope in a Taub-NUT background \cite{Taub:1950ez,Newman:1963yy}.

Now consider the time integral of $D_B\widetilde{A}{}^B$ in the memory effect \eqref{memory}. It is an analogue of gravitational spin memory \cite{Pasterski:2015tva} and it is related to the subleading soft photon theorem; one can see this by rewriting $A_a=\int\dd u\,\dot{A}_a$ as a time integral of the news, whereupon the linear term in eq.~\eqref{memory} is a double time integral reminiscent of subleading soft factors: see \eg ref.~\cite{Mao:2017axa}.

\medskip
\noindent\textbf{Memory and duality.} Let us now focus on the nonlinear term of the gyromagnetic memory \eqref{memory}. Our first claim is that it can be seen as a canonical generator of electric-magnetic duality transformations that are suitably local on future null infinity. To prove this, we begin with a reminder: electric-magnetic duality is a manifest symmetry of the vacuum Maxwell equations; it mixes electric and magnetic fields through rotations of the electromagnetic field $\cF$ and its Hodge dual $\ast\cF$. These can be enhanced to a symmetry of the action by defining a second gauge potential $\cC$ constrained by the nonlocal condition $\dd\cC=\ast\dd\cA$, whereupon duality transformations become U(1) rotations of the pair $(\cA,\cC)$. Things simplify near null infinity, where one has $\cC_\mu={C_\mu}/{r}+\cO(r^{-2})$ in terms of an $r$-independent function $C_\mu$; the above constraint then reduces to a \textit{local} relation $\dot{C}_a=\eps_a{}^b\dot{A}_b=\dot{\widetilde{A}}_a$, which yields $C_a=\widetilde{A}_a$ up to integration functions. From this perspective, duality induces global U(1) rotations of the pair $(A_a,\tlA_a)$. Further details on the interplay of duality and asymptotic symmetries can be found in \cite{Hosseinzadeh:2018dkh,Freidel:2018fsk,Henneaux:2020nxi,Liu:2023qtr}.
 
Now, as far as radiative data at large distances is concerned, one can in fact enhance duality transformations into \textit{local} rotations on a celestial sphere. Indeed, given any smooth function $\veps(\bth)$ on $S^2$, the infinitesimal transformations
\begin{align}
    \de_\veps A_B=\veps(\bth) \tlA_B\,,\qquad \de_\veps \tlA_B=-\veps(\bth) A_B
\end{align}
preserve the radiative symplectic structure \cite{Ashtekar:1981bq,Lee:1990nz}
\begin{align}
\label{symplectic structure}
    \Gamma
    &=
    \int\dd u\,\dd^2\bth\,\sqrt{\gamma}\,
    \de \dot{A}^B\wedge \de A_B\,.
\end{align}
The corresponding Hamiltonian generator can be found through the standard procedure and reads
\begin{align}
\label{duch}
    Q[\veps]&= \int\dd^2\bth \sqrt{\gamma} \,\veps(\bth) \,D(\bth)\,,
\end{align}
with the local density \footnote{The choice to write $\Delta{\widetilde{A}}$ instead of just ${\widetilde{A}}$ in eq.~\eqref{denss} amounts to a choice of time-independent integration one-form in the definition of the dual gauge field. In fact, one can change this at will using electric and magnetic large gauge transformations. Our choice here is motivated by the matching between the density \eqref{denss} and the nonlinear part of the gyromagnetic memory \eqref{memory}.}
\begin{align}
\label{denss}
    D(\bth)&= \int_{-\infty}^{+\infty}\dd u\,\dot{A}^B(u,\bth)\,\Delta \widetilde{A}_B(u,\bth)\,.
\end{align}
This all reduces to standard duality transformations for $\veps(\bth)=\text{const}$. However, locality is crucial here, since one can then identify the density \eqref{denss} with the nonlinear part of the gyromagnetic memory \eqref{memory}. The latter is thus related to local electric-magnetic duality on the celestial sphere, as was to be shown.

\medskip
\noindent\textbf{Memory and optical helicity.}
Our second claim is that the nonlinear term in the memory \eqref{memory} measures optical helicity. To prove this, let us write the local density \eqref{denss} in terms of photonic Fock space operators. We start from the mode expansion of the radiative gauge field (with the conventions of \cite{Strominger:2017zoo}),
\begin{align}
 	A_a(u,\bn)
  &=
  -i\int_0^{+\infty}\frac{\dd\omega}{\sqrt{2\pi}}\big(e^{i\omega u}\,b^\dagger_{a}(\omega,\bn)-\hc\big)\,,
 \end{align}
where $b^\dagger_{a}(\omega,\bn)$ creates a transverse photon of frequency $\omega$, propagating along $\bn$ with polarization along the direction $a$. The celestial density \eqref{denss} then becomes
 \begin{align}
 \label{duchess}
 D(\bth)
 &=
 \int_{-\infty}^{+\infty}\dd u\,\dot{A}^a\tilde{A}_a
 =2i\eps^{ab}\int\dd\omega\,\omega\,b_a^\dagger \,b_b\,.
 \end{align}
The geometric interpretation of this quantity is clearest in the helicity basis, \ie in terms of a complex null dyad $\hat{E}_a{}^B$ on $S^2$ such that the inverse volume form reads $\eps^{ab}=\hat{E}^a_{\;A}\hat{E}^b_{\;B} \,\eps^{AB}=\text{diag}(-i,i)$. Indeed, eq.~\eqref{duchess} then reduces to
 \begin{align}\label{helicity}
 	D(\bth)&=2\int_0^\infty \dd\omega\,\omega\,(b_+^\dagger \,b_+-b_-^\dagger \,b_-),
 \end{align}
which is known as the \textit{optical helicity} \cite{cameron2012optical,Hosseinzadeh:2018dkh}: it measures the difference between the numbers of right-handed and left-handed photons emitted in the direction $\th^A$ on the future celestial sphere \footnote{The factor $\omega$ in the integrand of \eqref{helicity} is due to the use of a Lorentz-invariant measure in the mode expansion; a similar factor appears in the number operator.}. We have thus confirmed the second interpretation of the nonlinear term in the gyromagnetic memory \eqref{memory}.

At this point, it is clear that gyromagnetic memory is closely related to the electromagnetic radiative phase space. But this comes with a word of caution: it does \textit{not} imply that gyromagnetic memory is due to a transition between inequivalent vacua, as would be the case for the more standard (and leading) ``kick'' memory \cite{Bieri:2013hqa}. A similar disclaimer holds for gyroscopic gravitational memory \cite{Seraj:2021rxd,Seraj:2022qyt}. Of course, suitable subleading asymptotic symmetries may exist, that would identify gyromagnetic memory with a vacuum transition; but this is an open question that will not be addressed here.

\section{Gyromagnetic memory\texorpdfstring{\\from localized sources}{}}
\label{sesource}

We conclude this work with an estimate of the gyromagnetic effect produced by nonrelativistic localized sources. This is achieved in two different ways. First, we directly solve \eqref{solMax} for a point charge and exhibit the result for two kinds of oscillatory motion. Second, we consider more general, arbitrary localized sources in the multipolar expansion and find the leading nonrelativistic effect in the gyromagnetic memory. This paves the way for similar gravitational analyses in the post-Newtonian framework \cite{Blanchet:2013haa}.

\medskip
\noindent\textbf{Point-like sources.} Consider a particle with charge $q$ that follows some (accelerated) path $\bR(t)$ in space. The corresponding charge density is $\rho(\bx,t)=q\,\delta^{(3)}(\bx-\bR(t))$ and the current density is $\bJ(\bx,t)=q\,\bv(t)\,\delta^{(3)}(\bx-\bR(t))$, where $\bv=\dot\bR$ is the particle's velocity. As a result, the $\by$ integral of the electromagnetic field \eqref{solMax} is straightforward and one finds
\beq
\cA_\mu(\bx,t)
=
q\int\limits_{-\infty}^t\!\!\dd t'\,
v_{\mu}(t')\,
\frac{\delta(t'-t+|\bx-\bR(t')|)}{|\bx-\bR(t')|}
\label{s25}
\ee
where $v_0\equiv1$ and $v_i=v^i=\dot R^i$. It now remains to carry out the time integral, taking into account the nonlinear dependence of the delta function on the integration variable $t'$. This is most easily done near null infinity, where one may expand $t-|\bx-\bR(t')|=u+\bn\cdot\bR(t')+\cO(1/r)$ and the electromagnetic field \eqref{s25} gives the following special case of eq.~\eqref{aasp}:
\beq
\label{s35}
\cA_\mu(\bx,t)
=
\frac{q}{r}\frac{v_{\mu}(t_*)}{1-\bn\cdot\bv(t_*)}
+\cO(r^{-2})\,.
\ee
Here $t_*$ is the root of the transcendental equation
\beq
t'=u+\bn\cdot\bR(t')\,,
\label{t35}
\ee
enforced by the delta function in \eqref{s25} at leading order in $1/r$. It can be found by iterating the function $u+\bn\cdot\bR(...)$ infinitely many times, starting from the `seed' $u$:
\begin{align}
\label{iteration}
t_*
&=
u+\bn\cdot\bR\bigg(u+\bn\cdot\bR\Big(u+\bn\cdot\bR\big(\cdots(u)\cdots\big)\Big)\bigg)\,.
\end{align}
Note that this iteration is guaranteed to converge. Indeed, the root of eq.~\eqref{t35} is unique and stable by virtue of the fact that the particle's velocity is lower than the speed of light. One can then truncate the iteration \eqref{iteration} to a finite $n$-fold composition to obtain the nonrelativistic approximation of the electromagnetic field, incorporating the effects of the source up to order $|\bv|^n=(|\bv|/c)^n$.

Let us illustrate this with two examples of sources. Consider first a particle that oscillates along the $z$ axis, so that $\bR(t)=-R\cos(\Omega t)\be_z$ for some length $R$ and some frequency $\Omega$ such that $v\equiv\Omega R\ll1$. Eq.~\eqref{s35} then yields $\cA_x=\cA_y=0$ and
\beq
\cA_z
\sim
\frac{q}{r}\,v\Big[\sin(\Omega u)-v\cos(2\Omega u)\cos(\theta)+\cO(v^2)\Big]\,,
\ee
where $\theta$ is the standard azimuthal polar coordinate and we used the crudely approximate root $t_*\sim u-R\cos(\theta)\cos(\Omega u)$ of eq.~\eqref{t35}. In terms of electromagnetic boundary data, one finds $F_{ab}=0$ and $\dot A_a\widetilde A^a=0$, so there is no magnetic dipole precession. Note that this remains true at any order in the nonrelativistic expansion, because the only nonzero component of the gauge field at null infinity is $A_{\theta}$, which only depends on $u$ and $\theta$.

Let us now turn to a source that breaks parity symmetry, namely a charged particle that moves along a circle in the $(x,y)$ plane:
\beq
\bR(t)
=
\big(R\cos(\Omega t),R\sin(\Omega t),0\big)\,,
\label{s45}
\ee
again with $v\equiv\Omega R\ll1$. Now limiting ourselves only to the leading order in the nonrelativistic expansion, we find in Bondi coordinates that $\cA_u=\cA_r=q/r$ and
\begin{align}
\cA_{\theta}
&\sim
qv\,\sin(\phii-\Omega u)\cos\theta\,,\\
\cA_{\phii}
&\sim
qv\,\cos(\phii-\Omega u)\sin\theta\,.
\end{align}
It follows again that $F_{ab}=0$, but this time the integrand of optical helicity does \textit{not} vanish:
\beq
\label{s5}
\dot A_a\widetilde A^a
=
-q^2R^2\Omega^3\cos\theta
+\cO(|v|^3)\,.
\ee
As a result, the precession rate \eqref{omm} is nonzero. It depends on the azimuthal location of the test magnetic dipole with respect to the source, but the average of the precession rate over the whole celestial sphere vanishes. One may expect a similar behaviour for the gyroscopic gravitational memory produced by inspiralling binary systems, up to the fact that the angular distribution should start from the $\ell=2$ harmonic.

\medskip
\noindent\textbf{Multipolar expansion.} Let us now consider a generic compact source with characteristic speed $v$, small with respect to the speed of light $c$. For such sources, the multipolar expansion provides an efficient approximation scheme, since multipole moments of order $\ell$ in the radiation are suppressed by a factor $(v/c)^\ell$. We use Gaussian units and reinstate $c$ until the end of this work, as a bookkeeping parameter that controls the order of the nonrelativistic expansion.

To begin, we decompose the radiative field \eqref{arad} in terms of two scalar functions $\phi^{\pm}$ with definite parity:
\begin{align}
    A_a(u,\bn)&=D_a\phi^+(u,\bn)+\eps_a{}^b D_b \,\phi^-(u,\bn)\,.
\end{align}
The linear and nonlinear terms in the gyromagnetic precession rate \eqref{omm} then are 
\begin{align}
    D_a \widetilde{A}^a&=-D^2 \phi^{-}\,,
    \label{linear precession}\\
    \dot{A}_a \tilde{A}^a&=  -\left(D_a \dot{\phi}{}^{+} D^a \phi^{-}-D_a \dot{\phi}{}^{-} D^a \phi^{+}\right)\nonumber\\ 
    &\quad+\epsilon^{a b}\left(D_a \dot{\phi}{}^{+} D_b \phi^{+}+D_a \dot{\phi}{}^{-} D_b {\phi}^{-}\right)\,.
    \label{nopre}
\end{align}
The scalars $\phi^\pm$ can be expanded in symmetric trace-free (STF) harmonics on the sphere as $\phi^\pm(u,\bn)\equiv \phi^\pm_L(u)\, n_L$, where $L=(i_1i_2...i_{\ell})$ is a multi-index and the STF coefficients $\phi_L^\pm(u)$ are \textit{radiative multipole moments}. These can be derived through a multipole expansion of eq.~\eqref{solMax} \cite{Damour:1990gj}, which leads to 
\begin{align}
\label{rasou}
    \phi^+_L&=\frac{1}{c^\ell \ell \ell!}\pd_u^\ell q^+_L \,,
    \qquad
    \phi^-_L=\frac{1}{c^{\ell+1} (\ell+1)!}\pd_u^\ell q^-_L
\end{align}
where the \textit{source multipole moments} $q^\pm_L$ are explicitly given in \cite[eqs.\ (4.17)]{Damour:1990gj}. In the nonrelativistic limit, they take the simple form
\begin{align}
\label{sumo}
    q^+_L&=\int d^3x \,\rho\, \hat{x}_L+\cO(1/c^2)\,,\\
    q^-_L&=\int d^3x \,(\bx\times \bJ)_{\langle i}x_{L-1\rangle}+\cO(1/c^2)\,.
\end{align}
Note from \eqref{rasou}--\eqref{sumo} that the leading-order effect in the nonrelativistic limit is determined by $\phi_i^+=\frac{1}{c}\dot{p}_i$, where $p_i$ is the source's electric dipole moment. As a result, the nonlinear precession in \eqref{omm} is given at leading order by the third term in \eqref{nopre}:
\begin{align}
\label{nolili}
    \dot{A}_a \tilde{A}^a&=\frac{1}{c^2} \left(\ddot{\bp} \times \dot{\bp}\right) \cdot\bn+\cO(1/c^3)\,.
\end{align}
At the same time, the linear term term in the precession \eqref{omm} is determined by \eqref{linear precession}; this is given at leading order by the magnetic dipole $\phi_i^-=\frac{1}{2c^2}\dot{m}_i$, so that
\begin{align}\label{prelidi}
    D_a \widetilde{A}^a&=\frac{1}{c^2} \dot{\bm}\cdot \bn+\cO(1/c^3)\,.
\end{align}
It follows that the full gyromagnetic memory \eqref{memory} is \footnote{It is worth keeping track of dimensions in \eqref{PHINN}: the angle $\Phi$ is dimensionless, the gyromagnetic ratio $k$ is a charge per unit mass, $u$ is time, and Gaussian units mean that $4\pi\epsilon_0\equiv1$. The factors of $c$ in \eqref{PHINN} are fixed by these dimensional constraints.} 
\begin{align}
\label{PHINN}
    \Phi(\bn)
    &\sim
    \frac{k}{r^2c^3}
    \left[\Delta \bm
    -\frac{k}{2c^3}\int du\left(\ddot{\bp} \times \dot{\bp}\right)\right]\cdot \bn
\end{align}
at leading order in the nonrelativistic limit. As a consistency check, one can return to the example of the circular source \eqref{s45}, where $\bm=qR^2\bOmega$ and $\bp(t)=q\bR(t)$ so that \eqref{prelidi} vanishes while the nonlinear term \eqref{nolili} does not; eq.~\eqref{PHINN} then reproduces \eqref{s5}, now from an explicit multipolar nonrelativistic expansion.

In a more general situation where both pieces in \eqref{PHINN} are non-vanishing, one might expect the second term to be negligible with respect to the first one, owing to the relative factor $c^{-3}$. However, this is not so because the \textit{accumulation} due to the time integral of the nonlinear term results in an effect that can be comparable to the linear term. This can be verified by considering a point charge $q$ with mass $m$ moving on a quasi-circular orbit with radius $R(t)$ in an attractive Coulomb potential $-Q/r$. Due to its acceleration, the particle radiates and $R(t)$ shrinks in time. The detailed evolution equation satisfied by $R(t)$ is found by noting that the rate of change of total mechanical energy $E(t)=-Qq/2R$ (potential $+$ kinetic) is balanced by the radiated energy which is given by the Larmor formula $\dot{E}=-\frac{2}{3}q^2 a^2/c^3$, where $a$ is the acceleration of the particle. This implies that $\dot{R}=-\frac{4}{3} (\tfrac{q}{m})^2\frac{Q q}{ R^2 c^3}$, which can be used to compute both terms in \eqref{PHINN}. In fact, the second term is thus found to be $-3/8$ times the first term, and the total gyromagnetic memory is 
\begin{align}
   \Phi(\bn)
   &=
   \frac{5q}{16 m r^2 c^3}\, \Delta \bm \cdot \bn +\cO(c^{-4})\,.
\end{align}
This confirms that both linear and nonlinear pieces in \eqref{PHINN} have the same order of magnitude in the nonrelativistic limit, despite their seemingly different dependence on $c$.

\section{Conclusion}

This paper was devoted to a subleading electromagnetic memory effect that affects the orientation of magnetic dipoles; it mimics the gyroscopic gravitational memory effect of \cite{Seraj:2021rxd,Seraj:2022qyt}. Our main results are the precession rate \eqref{omm} and the ensuing memory \eqref{memory}, both of which are deeply related to the magnetic (parity-odd) structure of the electromagnetic radiative phase space.

A sharp difference between the electromagnetic and gravitational setups is the simplicity of the former. In particular, we provided explicit formulas [\eg \eqref{s5} and \eqref{PHINN}] for the precession rate of magnetic dipoles due to bounded sources of electromagnetic radiation. Carrying out the analogous computation for gravity will undoubtedly be illuminating, but it is also much harder. We hope to address this problem in the future.

A natural question is whether gyromagnetic memory and/or its gravitational version can be observed. Again, the electromagnetic setup has a clear advantage here, since it can be produced in highly controlled experiments while the detection of gravitational waves has only been possible for a few years \cite{LIGOScientific:2016aoc}. Note that electromagnetic effects sensitive to the helicity \eqref{helicity} are known to occur in Nature, \eg in magnetized plasmas (see \cite{Maleknejad:2023nyh} and references therein). One may thus expect gyromagnetic memory to be very much within the reach of current experiments.

\section*{Acknowledgements}

We thank Guillaume Faye, Azadeh Maleknejad and Marios Petropoulos for fruitful discussions on related subjects. The work of B.O.\ is supported by the European Union's Horizon 2020 research and innovation program under the Marie Sk{\l}odowska-Curie grant agreement No.\ 846244. A.S.\ is supported by a Royal Society University Research Fellowship.


%

\end{document}